# Development in Astronomy and Space Science in Africa

The development of astronomy and space science in Africa has grown significantly over the past few years. These advancements make the United Nations Sustainable Development Goals more achievable, and open up the possibility of new beneficial collaborations.

**Mirjana Pović[1,2], Michael Backes[3,4], Paul Baki[5], David Baratoux[6], Solomon Belay Tessema[1], Zouhair Benkhaldoun[7], Michael Bode[8], Nana A. Browne Klutse[9], Phil Charles[10], Kevin Govender[11,12], Ernst van Groningen[13], Edward Jurua[14], Alemiye Mamo[1,15], Sivuyile Manxoyi[12], Vanessa McBride[11,12], Jamal Mimouni[16], Takalani Nemaungani[17], Pheneas Nkundabakura[18], Bonaventure Okere[19], Somaya Saad[20], Prospery C. Simpemba[21], Tefera Walwa[22], and Abdissa Yilma[1]**

1. Ethiopian Space Science and Technology Institute, Addis Ababa, Ethiopia; 2. Instituto de Astrofísica de Andalucía (CSIC), Granada, Spain, email: mpovic@iaa.es; 3. Department of Physics, University of Namibia, Windhoek, Namibia; 4. Centre for Space Research, North-West University, Potchefstroom, South Africa; 5. Technical University of Kenya, Nairobi, Kenya; 6. French National Research Institute for Sustainable Development, Géosciences Environnement Toulouse, Université de Toulouse, Toulouse, France; 7. Oukaïmeden Observatory and the Cadi Ayyad University, Marrakech, Morocco; 8. Botswana International University of Science and Technology, Palapye, Botswana; 9. Ghana Space Science and Technology Institute, Accra, Ghana; 10. University of Southampton, Southampton, United Kingdom; 11. Office of Astronomy for Development (IAU), Cape Town, South Africa; 12. South African Astronomical Observatory, Cape Town, South Africa; 13. International Science Programme, Uppsala University, Uppsala, Sweden; 14. Mbarara University of Science and Technology, Mbarara, Uganda; 15. East African Regional Office of Astronomy for Development, ESSTI, Addis Ababa, Ethiopia; 16. Mentouri University, Constantine-1, LPMPS, Algeria; 17. Department of Science and Technology, Pretoria, South Africa; 18. University of Rwanda, Kigali, Rwanda; 19. West African Regional Office of Astronomy for Development, NASRDA, Centre for Basic Space Science, Nsukka, Nigeria; 20. National Research Institute of Astronomy & Geophysics, Cairo, Egypt; 21. Southern African Regional Office of Astronomy for Development, Copperbelt University, Kitwe, Zambia; and 22. Ethiopian Space Science Society, Addis Ababa, Ethiopia

Until recently, South Africa with the Southern African Astronomical Observatory (SAAO) and Hartebeesthoek Radio Astronomy Observatory (HartRAO), Namibia with the High Energy Stereoscopic System (HESS), and Morocco, Algeria and Egypt with their optical observatories, were almost the only astronomy references in Africa. Also, South Africa, Egypt, Nigeria and Algeria were the only four African countries with established satellite programmes. In recent years, many other countries began research activities in astronomy and space science (A&SS), starting with institutional development, human capacity development (HCD), scientific research and networking. The African Union (AU) took important steps in promoting the development of A&SS on a continental scale for improving some of the main socioeconomic and environmental challenges that Africa is facing, and for achieving United Nations Sustainable Development Goals (UN SDGs). This Comment aims to provide an overview of the current status and future prospects of A&SS in Africa.

**Institutional development**
In East and northeast Africa, Ethiopia, Kenya and Sudan have made the most significant achievements. In 2005 in Ethiopia, the Institute of Geophysics, Space Science and Astronomy was established as a national seismological, geomagnetic and geodetic observatory. The Ethiopian Space Science Society played a crucial role in establishing the Entoto Observatory and Research Centre in 2013, creating the first optical observatory in East and Central Africa with its two 1-m telescopes (Fig. 1). In 2016, the Ethiopian Space Science and Technology Institute (ESSTI) was established with the aim of developing space science and technology in the country and wider region, and using it for addressing some of the main challenges that the country is facing, such as access to water, agricultural productivity and so on.

Ethiopia has serious plans to develop both space and ground-based systems related to satellite technologies. ESSTI is now working on its first microsatellite (~ 65 kg) and is planning to launch its first remote sensing and telecommunications satellites in the next few years. In addition, several public universities are becoming more engaged in A&SS. There are plans to establish a larger optical observatory in the near future in northern Ethiopia (Lalibela), with a ~ 4 m telescope, and to convert Ethiopia into an attractive location for future astronomical projects. In Kenya, a few projects underscore recent institutional developments. It became one of the Square Kilometre Array (SKA) partner countries (the only one in the East African region). A precursor to the SKA, the 32-m Longonot telecommunications dish is expected to be available this year for conversion into a radio telescope under the African Very Long Baseline Interferometry Network (AVN). Kenya has plans to establish its first optical observatory in the near future. Funding has been secured in 2018 through the UK Astronomy Technology Centre to complete surveys of two sites. In February 2017, the Kenya Space Agency was established with the aim to develop space science and technology, and the first Kenyan remote sensing CubeSat, developed in collaboration with the Japan Aerospace Exploration Agency, was launched in May 2018. In 2013, Sudan established the Institute of Space Research and Aerospace (ISRA) for developing different fields of space science and aerospace technology. Currently some of the main objectives of ISRA are the establishment of a satellite ground station, launch of an educational CubeSat, design of aerial surveillance systems and the establishment of an astronomical exploration centre and ground-based telescope. In addition, and in relation to space science, Sudan's National Centre for Research established the Institute of Remote Sensing and Seismology.

In West Africa, Ghana, Nigeria and Burkina Faso have made significant developments in A&SS. The Ghana Space Science and Technology Institute was established in 2012 as a coordinating body of A&SS development. Ghana is the only West African country to have joined the SKA as a partner country. As part of the AVN, the Ghana Radio Astronomy Observatory was established in August 2017, covering the frequency range 4–8 GHz with a single 32-m dish converted from a satellite communications Earth station antenna (Fig. 2). In 2017, Ghana launched its first remote sensing CubeSat to be used mainly for coastline long-term monitoring and mapping, and for HCD. In Nigeria, the Centre for Basic Space Science (CBSS) was created in 2001. A team of engineers and scientists at the CBSS successfully assembled and installed a 3-m radio telescope, in order to better understand the operation and control of such a telescope. CBSS is part of the National Space Research and Development Agency (NASRDA), established in 1998. NASRDA is the main body for implementing Nigeria's space programme, which includes the development of basic space science and technology, remote sensing, satellite meteorology, communication and information technology, and defence and security [1]. Since 2003, Nigeria has had three satellites in orbit including NigComSat-1R, built in China, which was launched in 2011 and boosted Internet and telecommunication services across the country. Currently, the NigeriaSat-2 Earth observation satellite is producing higher resolution images than any other UK-built satellite. In Burkina Faso, at the Université de Ouagadougou, a project is under development to establish an astronomical observatory for research on Mount Djogari and install the 1-m MarLy optical telescope that was moved from the La Silla Observatory in Chile to Burkina Faso in 2010 [2]. Senegal also has plans to develop research astronomy in the near future. The upcoming NASA experiment to observe a stellar occultation by the trans-Neptunian object 2014 MU69 is supported by the Senegalese government and will foster USA–Senegal collaborations in astronomy.

In North Africa, Egypt's National Research Institute of Astronomy and Geophysics (NRIAG) was established more than 110 years ago and conducts activities under five different departments, including A&SS. NRIAG also established the National Data Center, the Kottamia Center of Scientific Excellence in Astronomy and Space Sciences, and operates the Kottamia Astronomical Observatory (KAO), a solar telescope and a satellite laser-ranging station at Helwan, and Abu Simble and Misallat magnetic observatories. KAO is one of the oldest observatories in Africa, with its 75-inch telescope. The country has plans to build the largest telescope in the Arab region, the ~ 6-m Egyptian Large Optical Telescope, possibly in northern Sinai (site testing is in progress). Egypt's National Authority for Remote Sensing and Space Sciences has several different centres and laboratories, which collectively have launched six satellites between 1998 and 2014, and a new remote sensing satellite is on the way. The Space Weather Monitoring Center was established in 2008, aiming to support the Egyptian space programme, and help with space weather prediction. The Egyptian Space Agency is on the way to being established, and has already been voted on in Parliament. This agency would aim to work on the future development of space science and technology, create a satellite-manufacturing centre, launch satellites from Egyptian territory, and serve the country's strategy in areas of development and national security. In Algeria, most astronomy research activities are carried out at the Centre de Recherche en Astronomie Astrophysique et Géophysique. Apart from the Bouzaréah Observatory (and an 80-cm telescope), there are plans to build two new observatories: one in the Aurès (northern Algeria), and another near Tamanrasset (southern Sahara). The Observatoire National des Aurès will cooperate with the European Virgo programme to optically follow up gravitational wave detections. Space research is mainly undertaken at the Centre des Techniques Spatiales (CTS) at Arzew. The CTS is in charge of developing the Algerian microsatellite programme (AlSat series), which launched several satellites under the Algerian Space Agency. Two centres for the use and exploitation of satellite data were established and the Institute National de Géodésie et de Télédétection is the main remote sensing centre [3]. In Morocco, astronomy has developed significantly over the past few years, especially due to the inauguration of the Oukaïmeden Observatory in 2007. Different projects and collaborations are running, including: a remote 0.5-m telescope for a survey of small Solar System bodies called the Morocco Oukaïmeden Sky Survey (with Switzerland and France), the 0.6-m TRAPPIST-North telescope for detecting transiting planets and planetesimals (withBelgium), the RENOIR experiment for monitoring the ionosphere for space weather (with USA), and the OWL-Net 0.5-m telescope for satellite re-entry and the monitoring of near-Earth objects (with South Korea) [4]. The High Energy Physics and Astrophysics Laboratory also plays an important role in Moroccan astrophysical development. Morocco has just launched a satellite called Mohammed VI-A, part of the space science programme led by the Royal Centre for Remote Sensing. Tunisia does not yet have its own professional infrastructures in astronomy, but has set up a Tunisian Astronomical Society.

In southern Africa, South Africa (SA) continues to be the major player in A&SS. It is the only country in the south with an established national space agency (SANSA), and has launched two satellites. The Department of Science and Technology identified astronomy as one of the key drivers of their international science collaboration agenda, and developed the [National Strategy for Multiwavelength Astronomy](#) in 2015. The first phase of SKA-SA, MeerKAT, involving the construction of the 64-dish radio array, has now been completed (see the [Mission Control by Fernando Camilo](#)). The SKA-SA and HartRAO recently merged to become the South African Radio Astronomy Observatory. The SAAO continues operating the South African Large Telescope and some 20 other South African and international telescopes (as well as geophysical and space science instruments). SA is now hosting MeerLICHT (together with the Netherlands and UK), a robotic 0.65-m optical telescope that is synchronized with MeerKAT to scan the southern skies at optical and radio wavelengths simultaneously. The AVN is making progress, and all African SKA member countries (including Namibia, Botswana, Mozambique, Zambia, Mauritius and Madagascar) signed a memorandum of understanding in 2017 to collaborate on radio astronomy. It is planned that in phase 2 of the SKA (expected to begin in 2025), thousands of dishes will be built in SA and African partner countries. Mauritius (in collaboration with India) has operated a radio telescope (MRT) at 151.5 MHz since 1992. In Namibia, the University of Namibia (UNAM) and the Namibia University of Science and Technology (NUST) are the two institutes most involved in A&SS. UNAM is one of the founding members of HESS, a system of five imaging atmospheric Cherenkov telescopes, and one of the leading gamma-ray observatories in the world (see the [Mission Control by Mathieu de Naurois](#)). The Namibian Institute of Space Technology was established under the NUST. There are plans to build the first millimetre-wave radio telescope in Africa in Namibia (together with the Netherlands) [5]. Botswana has very ambitious plans for the next five years: completion of its AVN node and significant involvement in the SA HIRAX radio astronomy project; the establishment of a National Remote Sensing Centre; and the building of a National Optical Observatory. Zambia is in a process of converting the redundant 30-m Mwembeshi telecommunications Earth station into a radio telescope to form part of the AVN.

**Human capacity development**
The progress that African countries have made in HCD in recent years, often in collaboration with international institutions, is remarkable. The number of employed researchers in A&SS has increased dramatically. Many countries established graduate programmes at their universities (for example, South Africa, Egypt, Morocco, Ethiopia, Uganda, Namibia, Kenya, Sudan and Nigeria), while many others are in the process of doing so. Public awareness and outreach has increased exponentially in almost all countries. And at the same time networking among African institutions and between African and international institutions improved significantly, contributing to both HCD and research.
The Development in Africa with Radio Astronomy (DARA) and SKA-HCD projects are working on the AVN training programmes in all eight African SKA partner countries, and are contributing strongly to HCD in radio astronomy (see the Comment by Melvin Hoare).
Uppsala University's International Science Programme (ISP) provides longterm support (up to 20 years or more) and assists low-income countries to build and strengthen their domestic capacity for research and postgraduate education. In East Africa, ISP supports the East African Astronomy Research Network, comprising Muni University (the coordinating office) and Mbarara University of Science and Technology (MUST) in Uganda, the University of Rwanda, and the ESSTI in Ethiopia. MUST and the University of Rwanda (through the Rwanda Astrophysics, Space and Climate Science Research Group) are also supported as individual research groups.
In 2017 a new widely supported initiative named the Africa Initiative for Planetary and Space Sciences was launched, with its main objectives to: (1) connect African planetary and space science (PSS) researchers with their international peers by facilitating cross-border collaborations while overcoming traditional language barriers; (2) build a road map for PSS in Africa by identifying key research areas where scientists can make significant contributions; and (3) contribute to sustainable development in Africa through research, education and public outreach in PSS [6].
Since 2013, many projects across the continent related to HCD and networking have been supported by the International Astronomical Union's Office of Astronomy for Development (see the Comment by Vanessa McBride et al.). And from 2000 the South African National Astrophysics and Space Science Programme has trained hundreds of MSc students in A&SS from a variety of African countries. This includes many non-SA students who returned to their home countries and have greatly assisted in the creation of an African network for A&SS.

**Continental initiatives**
Recently, the AU developed the Common Africa Position on the Post-2015 Development Agenda to define a framework facilitating the achievement of the UN SDGs. The AU recognizes the importance of investing in space sciences and technology, and defines it as one of fundamental factors for achieving sustainable socioeconomic and environmental development of African countries. The AU recently published the African Space Strategy to create a space programme on a continental level, defining key priority areas: social, economic and political affairs and their union toward the total integration. To deliver on these areas, four fields of science were selected to be developed within the continent: Earth observations (to address Africa's socioeconomic opportunities and challenges), navigation and positioning (for improving safety-of-life applications), satellite communications (for improving information communication technologies for commercial purposes and for the broader public good, especially in rural areas), and space science and astronomy (for stimulating human capital and technological spin-offs). To help achieve these goals, the African Space Agency was recently established.

**Improving the future**
Africa has amazing potential due to natural (such as large areas with dark night skies) and human resources for scientific research in A&SS. At the same time, the continent is still facing many difficulties, and countries are now recognizing the importance of astronomy, space science and satellite technology for improving some of their main socioeconomic and ecological challenges. As described above, many achievements have already been realized due to incredible efforts, but much still remains to be done. Strengthening the networks between African countries is fundamental for achieving the proposed continental goals. Exchange of knowledge and experience with international institutions and research groups, and the promotion of new collaborations are very much needed at this moment and can be beneficial for all. Only in this way, working together on the development of science and education in Africa, can we fight poverty in the long term and increase our possibilities of attaining the UN SDGs in future.

**Acknowledgements**
This paper is dedicated to all of the people who somehow contributed to the development of A&SS in Africa. Without them all of this would not be possible. In addition, this paper was inspired by sessions SS23 and LS7 during the 2018 European Week of A&SS (EWASS). Both sessions were supported by the UK Science and Technology Facilities Council, UK Royal Astronomical Society, International Astronomical Union Office of Astronomy for Development, European Astronomical Society, International Science Programme, and Development in Africa with Radio Astronomy project.

**References**
1. Boroffice, R. A. Afr. Skies 12, 40–45 (2008).
2. Carignan, C. Afr. Skies 16, 18–20 (2012).
3. Mimouni, J. in The Role of Astronomy in Society and Culture (IAU Symposium 260) 741–747 (Cambridge Univ. Press, Cambridge, 2011).
4. Benkhaldoun, Z. Nat. Astron. 2, 352–354 (2018).
5. Backes, M. et al. Proc. Sci. https://doi.org/10.22323/1.275.0029 (2017).
6. Baratoux, D. et al. Eos https://doi.org/10.1029/2017EO075833 (2017).

**Figure 1. [Image not included here due to Nature Astronomy copyright restrictions. Link to published version of paper is provided in the ArXiv 'Comments' field]** The Entoto Observatory and Research Centre, Addis Ababa, Ethiopia. a) The observatory; b) The 1-m telescope. Credit: Getnet Gebereegziabher

**Figure 2. [Image not included here due to Nature Astronomy copyright restrictions. Link to published version of paper is provided in the ArXiv 'Comments' field]** The Ghana Radio Astronomy Observatory, Kuntunse. The Ghanaian engineers were trained by South African experts. Credit: Ghana Space Science and Technology Institute